\documentclass[10pt,times]{iopart}
\usepackage{epsfig}
\usepackage{graphics}
\usepackage{graphicx}
\usepackage{bm}
\usepackage{iopams}
\usepackage{color}

\begin{document}


\newcommand{\dt}{\mathrm{d}t}
\newcommand{\dx}{\mathrm{d}x}
\newcommand{\geo}{GEO\,600}

\noindent {\small AEI-2006-019}

\title[Null-stream veto]{Null-stream veto for two co-located detectors:
Implementation issues}

\author{P Ajith\dag, M Hewitson\dag ~and I S Heng\ddag}

\address{\dag~Max-Planck-Institut f\"ur Gravitationsphysik (Albert-Einstein-Institut) \\
und Universit\"at Hannover, Callinstr.~38, 30167 Hannover, Germany}
\address{\ddag~Department of Physics and Astronomy, University of Glasgow, \\
Glasgow, G12 8QQ, United Kingdom}

\eads{\mailto{Ajith.Parameswaran@aei.mpg.de}}

\date{\today}

%

\begin{abstract}

Time-series data from multiple gravitational wave (GW) detectors can be linearly
combined to form a \emph{null-stream}, in which all GW information will be cancelled
out. This null-stream can be used to distinguish between actual GW triggers and 
spurious noise transients in a search for GW bursts using a network of detectors. 
The biggest source of error in the null-stream analysis comes from the fact that 
the detector data are {\it not} perfectly calibrated. In this paper, we present 
an implementation of the null-stream veto in the simplest network of two co-located
detectors. The detectors are assumed to have calibration uncertainties and correlated
noise components. We estimate the effect of calibration uncertainties in the null-stream
veto analysis and propose a new formulation to overcome this. This new formulation 
is demonstrated by doing software injections in Gaussian noise.   

\end{abstract}


\section{Introduction}

Given the time series data from a network of gravitational-wave (GW) detectors, one 
can find a particular linear combination of the data streams such that it does not 
contain any trace of GWs. The idea of this {\it null-stream} was proposed by G\"ursel and 
Tinto in their classic work \cite{GT}. G\"ursel and Tinto proposed that the null-stream
can be used to solve the `inverse problem' of GW bursts, i.e., to compute the unknown
quantities (two sky-positions and two polarizations) associated with the gravitational-waveform
from the responses of three broad-band detectors. 

Recently, there has been a lot of interest in the null-stream among the GW community. 
The main reason for this rejuvenated interest is that the first generation of ground-based 
interferometric GW detectors~\cite{ligo,virgo,geo,tama} have started acquiring scientifically 
interesting data. Among the most promising astrophysical sources of GWs for these ground-based 
detectors are the transient, unmodelled astrophysical phenomena like supernovae explosions, 
Gamma-ray bursts and black hole/neutron star mergers - popularly known as `unmodelled bursts'.  
Most of the algorithms currently used in burst-searches are time-frequency detection 
algorithms that look for short-lived excitations of power in the `time-frequency map' 
constructed from the data~\cite{ExPower,TFClusters,Waveburst}. Since present-day interferometric 
GW detectors are highly complex instruments, the data often contains lots of noise transients 
which trigger the burst detection algorithms. It is almost impossible to distinguish these 
spurious instrumental bursts from actual GW bursts using any physical model of the GW 
bursts. Thus, burst-data-analysis is usually performed as a coincidence analysis between 
multiple detectors. Although this `coincidence requirement' considerably reduces the 
list of candidate burst triggers, one month of data can potentially produce hundreds of 
multi-detector random coincidences. So, it becomes absolutely necessary to have 
additional `waveform consistency tests' that distinguish actual GW bursts from spurious
noise transients. A cross-correlation statistic that is formulated in~\cite{rStatistic} is
already being used as a coherent waveform consistency test in the search for GW bursts in the data  
of LIGO detectors. This is making use of the fact that all the LIGO detectors are
aligned (approximately) parallel to each other.   
 
Recently, it was proposed by~\cite{WS} that the null-stream can be used to distinguish 
between actual GW triggers and spurious noise transients in a search for GW bursts 
using any general network of detectors. The main idea is that, if the coincident triggers 
correspond to an actual GW burst, the null-stream constructed at the time of the 
triggers will contain no trace of the burst, and, will fall in to an expected 
noise-distribution. On the other hand, if the coincident triggers correspond 
to spurious instrumental bursts, the bursts will not necessarily cancel out in the null-stream, 
and the null-stream will contain some excess power. Many authors have proposed similar,
but non-equivalent ways of implementing this. Our method is based on the {\it excess power}
statistic~\cite{ExPower}, which was first used by~\cite{WS} in this context. This 
method is formally developed and is studied in detail in~\cite{NSMethod}. 
For an alternative implementation, see~\cite{CIT-JPL}.
A similar veto strategy using the null-stream constructed from the two 
calibrated output quadratures of \geo\ \cite{geo} detector is already being used to veto 
the burst triggers from \geo. This is discussed in~\cite{hNull}.

It was soon realised that the biggest source of error in the null-stream analysis 
comes from the fact that the detector data are {\it not} perfectly calibrated, due
to various reasons. In such cases, the null-stream constructed from the data 
containing actual GW bursts will contain some residual signal and will deviate from 
the expected noise distribution. This paper tries to address such practical issues 
connected with the implementation of the null-stream veto in the burst-data-analysis 
using a network consisting of two co-located interferometric detectors, like the two 
LIGO detectors~\cite{ligo} in Hanford, WA. The detectors are allowed to have calibration 
uncertainties 
and correlated noise components. In Section~\ref{sec:NSveto}, we briefly review the 
veto method in the case of two co-located detectors. In Section~\ref{sec:CalErr}, we 
estimate the effect of calibration uncertainties in the null-stream veto, and in 
Section~\ref{sec:CalErr}, we lay out and demonstrate a formulation to overcome this effect.


\section{The null-stream veto}
\label{sec:NSveto}
In the case of detectors placed widely apart, the null-stream is a function of the 
antenna patterns, and hence, the source-position~\cite{GT}. But in the case of two 
co-located detectors, the null stream is particularly simple. If $h_1(t)$ and $h_2(t)$
denote the properly calibrated time series data from the two detectors, the null-stream
is just~\cite{WS}: 
\begin{equation}
n(t) = h_1(t) - h_2(t),
\label{eq:nullstream}
\end{equation}
or, in discrete notation
\begin{equation}
n_j = h_{1_j}- h_{2_j}, 
\end{equation}
where $h_{1_j}$ and $h_{2_j}$ are the sampled versions of $h_1(t)$ and $h_2(t)$.  At the 
time of a set of coincident triggers, we construct the null-stream $n_j$. Let $\tilde N_k$ 
denote the discrete Fourier transform (DFT) of $n_j$ computed using $L$ samples of the 
data. We assume that the real and imaginary parts of $\tilde N_k$ are derived from a 
multivariate Gaussian distribution of mean zero~\footnote{If the real and imaginary parts 
of $\tilde N_k$ are derived from a non-zero-mean Gaussian distribution, one can always 
convert them to mean-zero Gaussian variables by subtracting the sample mean $\mu_k$ from them.} 
and variance $\sigma^2_k$. Following~\cite{ExPower,WS}, we compute the `excess power' 
statistic from the null-stream: 
\begin{equation}
\epsilon = \sum_{k=m}^{m+M} P_k \,,\, P_k = \frac{|\tilde N_k|^2}{\sigma_k^2}. 
\label{eq:stat}
\end{equation}
It can be shown that $\epsilon$ will follow a $\chi^2$ distribution of $2M$ degrees of 
freedom in the case of a non-windowed DFT. But in the case of a windowed DFT, $P_k$ are 
not {\it independent} $\chi^2$ 
variables, and hence $\epsilon$ will {\it not} follow a $\chi^2$ distribution~\cite{JL}. 
But,we note that the $\chi^2$ distribution is a special case of the Gamma distribution. It 
can be shown that, to a very good approximation, $\epsilon$ will follow a Gamma distribution 
with scale parameter $\alpha$ and shape parameter $\beta$. These parameters are related 
to the mean $\mu_\epsilon$ and variance $\sigma^2_\epsilon$ of the distribution of $\epsilon$ 
by
\begin{equation}
\alpha = \Big(\frac{\mu_\epsilon}{\sigma_\epsilon}\Big)^2, \,\,\,\, 
\beta = \frac{\sigma^2_\epsilon}{\mu_\epsilon}.
\end{equation} 
In order to estimate the parameters of the expected Gamma distribution, we generate a population of 
$\epsilon$ from stationary data (i.e., data not containing the burst event under investigation, 
but surrounding it). To be explicit, we divide the data in to a number of segments each length
$L$ and compute $\epsilon$ from each of these segments. From that population, $\mu_\epsilon$ 
and $\sigma^2_\epsilon$ can be estimated, and hence $\alpha$ and $\beta$.

It is known that the maximum signal-to-noise ratio (SNR) for the excess power statistic 
is achieved when the time-frequency volume used to compute the statistic is equal to the 
actual time-frequency volume of the signal~\cite{ExPower}. Since the duration and bandwidth 
of the burst is estimated by the burst detection algorithm itself, this information is 
used to decide on the length ($L$) of the data used to compute $\tilde N_k$ and the bandwidth 
over which $P_k$ is summed over. 

If the $\epsilon$ computed from the segment containing the burst is greater than a threshold, 
we veto the trigger. The threshold, $\tau$, giving a false-dismissal (`false-veto') probability 
of $\gamma$ can be found by 
\begin{equation}
\gamma = \int_\tau^\infty f(x;\alpha,\beta) \, \dx,
\label{eq:thresh}
\end{equation}
where $f(x;\alpha,\beta)$ is the probability density of the Gamma distribution with 
parameters $\alpha$ and $\beta$.

\subsection{Software injections}
\label{sec:SWinj}
Let us define some terminology. The false-dismissal probability is the {\it probability} 
of an actual GW burst being falsely vetoed, and the `false-veto fraction' is the fraction of
GW bursts that are {\it actually} vetoed using this method. As a sanity check, we estimate 
the false-veto fraction by injecting some prototype gravitational-waveforms into two data 
streams of Gaussian white-noise and performing the analysis. If all of our 
assumptions are true, the fraction of vetoed events among the injections should be equal 
to the chosen false-dismissal probability. 

The injected waveforms are Gaussian-modulated sinusoidal waveforms, of the form:
\begin{equation}
\hat h(t) = \hat h_{\rm rss}\left(\frac{2 f_0^2}{\pi}\right)^{1/4} \, \sin\left[2\pi f_0(t-t_0)\right] \, 
\exp\left[-(t-t_0)^2/\tau^2\right] \,,
\end{equation}
where $f_0$ is the central frequency of the waveform (randomly chosen from the 
set \{153, 235, 361, 554, 849, 1053, 1245, 1534, 1856\} Hz) and $t_0$ is the time corresponding 
to the peak amplitude. We setup the envelope width as $\tau = 2/f_0$, which gives durations of
approximately 1-20 ms. The corresponding quality factor is $Q \equiv \sqrt{2}\pi f_0\tau = 8.9$
and bandwidth is $\Delta f = f_0/Q \simeq 0.1 f_0$. The quantity $\hat h_{\rm rss}$ is 
the root-sum-squared (RSS) amplitude:
\begin{equation}
\left[\int_{-\infty}^{\infty} \hat h^2(t) \, \dt \, \right]^{1/2} = \hat h_{\rm rss} \, .
\end{equation}
The $\hat h_{\rm rss}$ is randomly chosen from the logarithmically-spaced interval 
$(5\times10^{-22},1\times10^{-19})$.
The amplitude spectral density (ASD) of the noise in the two data streams is chosen to be 
$1\times10^{-22}/\sqrt{\rm Hz}$ and $2\times10^{-22}/\sqrt{\rm Hz}$. We define the combined 
SNR, $\rho$, by 
\begin{equation}
\rho^2 = \rho_1^2+\rho_2^2 \,,
\label{eq:snrdefn}
\end{equation}
where $\rho_1$ and $\rho_2$ are the optimal SNRs~\footnote{The optimal SNR $\rho$ in 
detecting a signal $h(t)$ buried in the noise is defined by 
$\rho^2 = 4 \int_0^\infty |\tilde{H}(f)|^2 df / S_n(f)$ where $\tilde{H}(f)$ is the 
Fourier transform of the signal and $S_n(f)$ is the one-sided PSD of the detector noise.} 
in detecting the bursts in the two data streams, and quote this quantity while discussing 
the results.

The veto analysis is performed with different thresholds. The fraction of 
vetoed events is plotted against the false-dismissal probability corresponding to the 
chosen threshold in Figure~\ref{fig:FalseVeto} (left). It can be seen that the estimated 
false-veto fraction is in very good agreement with the predicted false-dismissal probability.


\section{Calibration uncertainties}
\label{sec:CalErr}

\begin{figure}[tb]
\centering
\includegraphics[width=12.5cm]{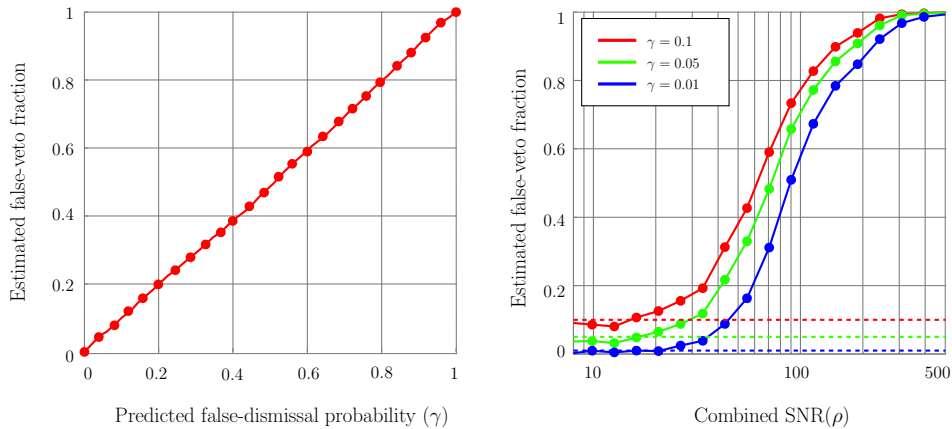}
\caption {[{\it Left}]: Estimated false-veto fraction plotted against the predicted
false-dismissal probability, assuming that the two detectors are perfectly calibrated. 
[{\it Right}]: Estimated false-veto fraction in the presence of $\pm 10\%$ calibration 
uncertainty, plotted against the combined SNR of the injections, for three different values 
of the predicted false-dismissal probability ($\gamma$). The three dashed lines  
show the predicted values of the false-dismissal probability in the absence of calibration 
uncertainties.} 
\label{fig:FalseVeto}
\end{figure}
%
So far, we have been assuming that the two data streams are perfectly calibrated. 
But, due to various limitations in the calibration procedure, the calibration of 
present-day interferometers can be subject to uncertainties of a few percent in 
amplitude and a few degrees in phase. This means that the null-stream constructed from 
the data containing actual GW triggers can contain some residual signal, and the 
computed statistic can vary from the expected distribution. This will result in a 
different false-dismissal probability than the one predicted by the hypothesis test.

It can be shown that we see approximately the same residual signal power in the 
null-stream for both a 10\% relative amplitude error and a 10 degree relative phase 
error; therefore, in principle, we need to consider both of these effects. However, 
it is not easy to conceive of a simple model for the possible relative phase error 
between two co-located detectors (since such errors most probably arise due to 
inaccuracies in the calibration process and will therefore be frequency dependent)
\footnote{One another possible source of relative phase error would be a time-offset 
between the two data streams. Although it is unlikely that such an error exists at 
any significant level, it would be possible to include this in the following analysis 
if necessary.}. It is, however, easy to think of a simple model for one possible 
source of relative amplitude calibration error. Suppose we know the absolute 
calibration of each detector to some accuracy. Then we have the possibility that 
the difference between the two data streams contains a systematic calibration error 
that would be independent of frequency. We explore a way to deal with this type of 
error in the rest of this paper.
   
As a simple model, we assume that the frequency-dependence of the calibration error
is negligible, and that the calibration error is a constant scaling factor over short 
time-scales (of the order of seconds). In the context of this analysis, we can assume 
that one detector is perfectly calibrated and the other is calibrated with a wrong 
scaling factor, i.e.,
\begin{eqnarray}
h_1(t) = n_1(t) + \hat h(t) \,, \nonumber \\ 
h_2(t) = n_2(t) + (1+\delta) \, \hat h(t) \,,
\label{eq:datawithcalerr}
\end{eqnarray}
where $n_1(t)$ and $n_2(t)$ are the detector noises in which a gravitational waveform $\hat h(t)$
is present. $\delta$ is the relative calibration error which is assumed to be a real quantity
and is constant over short time-scales. 

In order to estimate the effect of calibration error in the false-dismissal probability,
injections are done simulating a relative calibration error of $\delta = \pm 0.1$ between 
the two data streams. The fraction of the vetoed events among the injections is shown in 
the right plot of Figure~\ref{fig:FalseVeto} as a function of the combined SNR of the 
injections. Different curves in the plot represent three different values of the predicted 
false-dismissal probability ($\gamma$). It can be seen that the estimated false-veto fraction
raises to alarmingly high values for strong signals. 

\subsection{Dealing with calibration uncertainties}

\begin{figure}[tb]
\centering
\includegraphics[width=6.2cm]{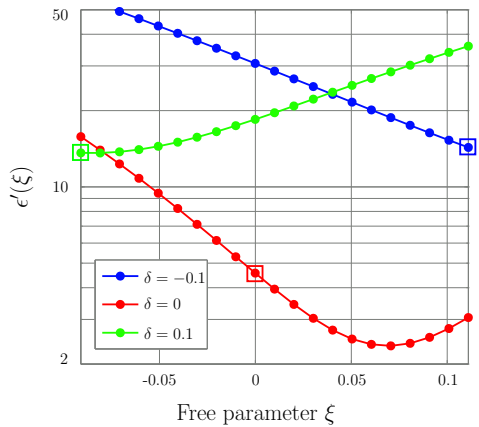}
\caption {The excess-power statistic $\epsilon'(\xi)$ constructed from null-stream $n'(t;\xi)$ 
plotted against the free parameter $\xi$, for three different values of $\delta$. We 
assume that the maximum possible absolute value of $\delta$ is 0.1. 
$\epsilon'\left(-\delta/(1+\delta)\right)$ for each curve is marked with a square. It can be 
seen that the minimum value of $\epsilon'(\xi)$ is less than, or equal to $\epsilon'(-\delta/(1+\delta))$.}
\label{fig:CalErrMinimizazion}
\end{figure}

In this section we formulate a strategy to reduce the effect of calibration errors in 
the false-dismissal probability. We construct the following linear combination of the 
data streams by introducing a free parameter, $\xi$, in the null-stream construction, i.e.,
\begin{equation}  
n'(t;\xi) = h_1(t) - (1+\xi) \, h_2(t) \,.
\label{eq:genNS}
\end{equation}
Substituting for $h_1(t)$ and $h_2(t)$ from Eq.(\ref{eq:datawithcalerr}) gives,
\begin{equation}  
n'(t;\xi) = n(t;\xi) - (\delta+\xi+\xi \, \delta) \, \hat h(t) \,,
\label{eq:genNS2}
\end{equation}
where 
\begin{eqnarray}
n(t;\xi) &=& n_1(t) - (1+\xi) \, n_2(t) \,,
\end{eqnarray}
is the `perfect' null-stream. We can `tune' the parameter $\xi$ such that the residual signal
disappears in Eq.(\ref{eq:genNS2}). This is accomplished by  minimizing the `excess power' in 
$n'(t;\xi)$ by varying $\xi$ over an interval.

As described in Section~\ref{sec:NSveto}, the null-stream $n(t;\xi)$ is divided into a number of 
short segments and the DFT of each segment is computed. It may be noted that in all segments
except the one containing the burst, $n'(t;\xi)=n(t;\xi)$, because the signal is absent in these
segments. We denote the DFT of $n(t;\xi)$ and 
$n'(t;\xi)$ by $\tilde N_k(\xi)$ and $\tilde N'_k(\xi)$, respectively. The mean, $\mu_k(\xi)$, 
and variance, $\sigma^2_k(\xi)$, of $\tilde N_k(\xi)$ are estimated from the neighboring segments 
of the one containing the burst. These are the moments of the expected distribution of $\tilde N_k'(\xi)$ 
{\it in the absence of the burst}. $\tilde N_k(\xi)$ is converted to a mean-zero Gaussian variable
by subtracting the sample mean $\mu_k(\xi)$ from it. 
The test statistic $\epsilon'(\xi)$ is computed from the segment of $n'(t;\xi)$ containing the 
burst:
\begin{equation}
\epsilon'(\xi) = \sum_{k=m}^{m+M} P'_k(\xi) \,,\, P'_k = \frac{|\tilde N'_k(\xi)|^2}{\sigma_k^2(\xi)}. 
\label{eq:stat}
\end{equation}
We minimize $\epsilon'(\xi)$  by varying $\xi$ over an interval $(\xi_{\rm min},\xi_{\rm max})$
~\footnote{It is important to note that we minimise the \emph{excess-power}, $\epsilon'(\xi)$,  in 
the null-stream, and not the \emph{total power}, $n'^2(t;\xi)$. In the later case, if there 
are correlated noise in $h_1(t)$ and $h_2(t)$, $\xi$ can take values which will minimise the 
correlated noise components in $n'(t;\xi)$ instead of the residual signal.}. 
The boundary of the parameter-space can be fixed as
\begin{eqnarray}
\xi_{\rm min} = \frac{-\delta_{\rm max}}{1+\delta_{\rm max}} \,,~~ 
\xi_{\rm max} = \frac{\delta_{\rm max}}{1-\delta_{\rm max}}\,,
\end{eqnarray} 
were $\pm\delta_{\rm max}$ is the maximum allowed value of the calibration uncertainty.  
When $\xi \rightarrow -\delta/(1+\delta)$, the residual signal in the null-stream 
cancells out,  and $\epsilon'(\xi) \rightarrow \epsilon(\xi)$, where 
\begin{equation}
\epsilon(\xi) = \sum_{k=m}^{m+M} P_k(\xi) \,,\, P_k = \frac{|\tilde N_k(\xi)|^2}{\sigma_k^2(\xi)}, 
\label{eq:stat}
\end{equation}
which falls in to an expected Gamma distribution in the case of GW bursts.

Since $\epsilon$ is quadratic in $\tilde N'_k$, apart from $|\tilde N_k|^2$ and 
$|(\delta+\xi+\xi\,\delta)\,\hat H_k|^2$, it contains also the cross-terms. 
This means that the minimum value of $\epsilon'(\xi)$ could be less than $\epsilon(\xi)$.
This is illustrated in Figure~\ref{fig:CalErrMinimizazion}, for three different values of $\delta$. 
Thus, the obtained false-dismissal probability could be less than what is predicted by the 
hypothesis test. Since the cross-terms depend upon the actual value of $\delta$, this
adds an error-bar to the false-dismissal probability. {\it But it may be noted that the 
actual false-dismissal probability is always less than (or equal to) what is predicted by the 
hypothesis test.}  

\subsection{Software injections with simulated calibration errors}

\begin{figure}[tb]
\centering
\includegraphics[width=12.5cm]{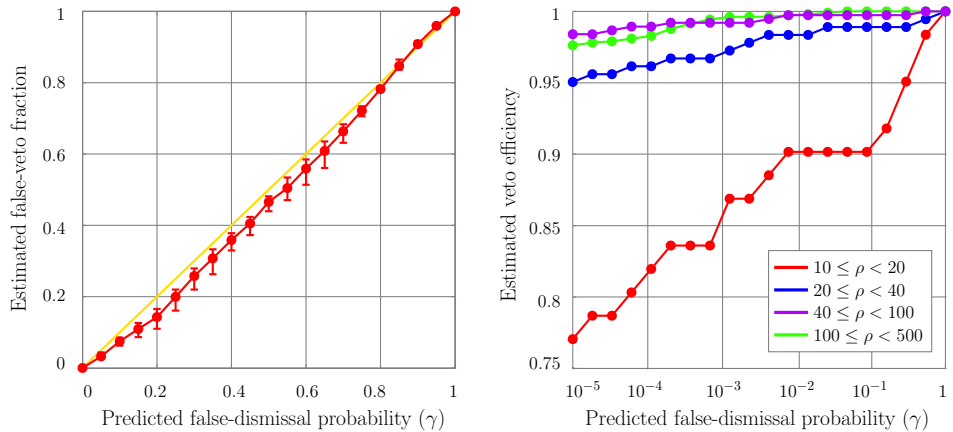}
\caption {[{\it Left}]: Estimated false-veto fraction plotted against the predicted 
false-dismissal probability. The relative calibration between $h_1(t)$ and $h_2(t)$
is assumed to be between $-0.1 \leq \delta \leq 0.1$.  
[{\it Right}]: Rejection power of the veto plotted against the predicted false-dismissal 
probability. Each curve in the plot corresponds to a particular range of SNR.}
\label{fig:WN_FalseVeto_Eff}
\end{figure}

We generate two data streams with a simulated relative calibration error $\delta$ 
according to Eq.(\ref{eq:datawithcalerr}) and perform the veto analysis, assuming 
that the $|\delta_{\rm max}| = 0.1$. The minimisation of $\epsilon(\xi)$ is carried out 
using an optimised minimisation algorithm. The analysis is performed for three different 
values (-0.1, 0, 0.1) of $\delta$ and the false-veto fractions corresponding to different 
thresholds are estimated in each case (in all cases, we assumed that $|\delta_{\rm max}| = 0.1$). 
The mean value (of the three simulations) of the false-veto fraction corresponding 
to each threshold is plotted against the corresponding false-dismissal probability 
in Figure~\ref{fig:WN_FalseVeto_Eff} (left). The extremum values corresponding to 
each threshold are used to generate the error-bars. It can be seen that the estimated 
false-veto fraction is always less than or equal to the predicted false-dismissal 
probability. 

The real figure-of-merit of a veto method is its ability to reject spurious events 
with a given false-dismissal probability. But, given that the probability density 
of the noise transients are not known {\it a priori}, there is no rigorous way of 
estimating the `rejection power' of the veto. The best we can do is to estimate the 
ability of the veto to reject a given glitch population. As a plausible estimation, 
we inject a population of sine-Gaussian waveforms with random parameters into two 
data streams. We then perform the veto analysis after choosing different thresholds. 
The estimated rejection power is plotted against the false-dismissal probability in 
Figure~\ref{fig:WN_FalseVeto_Eff} (right). Since the `excess power' in the null-stream
is proportional to the individual SNRs of the bursts in the two data streams, the 
rejection power is also proportional to the SNR. Each curve in the figure corresponds
to a particular range of SNR $\rho$ (see Eq.(\ref{eq:snrdefn})). This suggests that veto 
efficiencies of $\geq$ 90 \% can be achieved with a false-dismissal probability of 
$\simeq$ 1\% for spurious noise transients with $\rho \geq 10$.

\section{Summary}

Th null-stream constructed from the data of multiple GW detectors can be used to distinguish 
between actual GW triggers and spurious noise transients in the search for GW bursts using a 
network of detectors. The biggest source of error in the analysis comes from the fact that 
the present-day detectors are subject to calibration uncertainties. In this paper we have 
proposed an implementation of the null-stream veto in the search for GW bursts in the data of 
two co-located interferometers. We have estimated the effect of calibration uncertainties in the veto 
analysis by performing software injections in Gaussian noise with simulated calibration errors. 
A strategy is proposed to minimize this effect, assuming a simple model for the amplitude 
calibration-error and neglecting the errors in the phase calibration. This is done by introducing 
an additional free parameter in the null-stream combination and minimizing the excess-power in 
the null-stream. Finally, we compared the estimated fraction of falsely-vetoed GW-like injections 
with the predicted false-dismissal probability and found that the estimated fraction has a good 
agreement with the prediction. We also estimate the rejection power of the veto as a function 
of the false-dismissal probability.

\section*{Acknowledgments}

The authors are grateful to Bruce Allen, Peter Shawhan, Bernard Schutz, 
Linqing Wen and members of the LIGO Scientific Collaboration's Working Group on Burst
Sources for useful discussions. 

\bigskip


\end{document}